\documentclass[a4paper,11pt]{article}
\usepackage{amsmath,amssymb,amsfonts,amsthm,latexsym} 

\usepackage{pos}
\usepackage[normalem]{ulem}
\usepackage[utf8]{inputenc}
\usepackage[T1]{fontenc}
\usepackage{cancel}
\usepackage{graphicx}
\usepackage{latexsym}
\usepackage{bbm}
\usepackage{epsfig}
\usepackage{epstopdf}
\usepackage{subcaption}
\usepackage{color}
\usepackage{comment}
\usepackage{slashed}
\usepackage{placeins}
\usepackage{cleveref}
\usepackage{setspace}
\usepackage{url}
\usepackage{tabularx}
\usepackage{xcolor}
\usepackage{enumitem}
\usepackage{hepunits}
\usepackage{hyperref}
\usepackage{braket}
\usepackage{diagbox}
\usepackage{dcolumn}
\usepackage{xfrac}
\usepackage{pgf}

\newcommand{\dd}[1]{\mathop{\mathrm{d}#1}}

\newcommand{\Ns}{N_{\rm s}}
\newcommand{\Nt}{N_{\rm t}}
\newcommand{\Tc}{T_{\rm c}}
\newcommand{\CF}{C_{\rm F}}

\newcommand{\GE}{G_\mathrm{E}}
\newcommand{\GB}{G_\mathrm{B}}
\newcommand{\GEB}{G_{\mathrm{E},\mathrm{B}}}

\newcommand{\kappaE}{\kappa_\mathrm{E}}
\newcommand{\kappaB}{\kappa_\mathrm{B}}
\newcommand{\kappaEB}{\kappa_{\mathrm{E},\mathrm{B}}}
\newcommand{\tauT}{\tau T}
\newcommand{\tauf}{\tau_\mathrm{F}}
\newcommand{\tauF}{\tau_\mathrm{F}} 

\newcommand{\rhoEB}{\rho_{\mathrm{E},\mathrm{B}}}
\newcommand{\rhoE}{\rho_{\mathrm{E}}}
\newcommand{\rhoB}{\rho_{\mathrm{B}}}

\newcommand{\ReTr}{\textrm{Re}\,\textrm{Tr}\,}
\newcommand{\MSb}{\overline{\textrm{MS}}}

\newcommand{\be}{\begin{equation}} 
\newcommand{\ee}{\end{equation}}
\def\ba#1\ea{\begin{align}#1\end{align}}
\newcommand{\bea}{\begin{eqnarray}} 
\newcommand{\eea}{\end{eqnarray}}
\def\lsim{\mathrel{\raise.3ex\hbox{$<$\kern-.75em\lower1ex\hbox{$\sim$}}}}
\def\gsim{\mathrel{\raise.3ex\hbox{$>$\kern-.75em\lower1ex\hbox{$\sim$}}}}

\renewcommand{\sfrac}[2]{\ensuremath{#1/#2}}
\newcommand{\TM}{{\sfrac{T}{M}}~}

\renewenvironment{thebibliography}[1]{
  \begin{oldthebibliography}{#1}
    \setlength{\itemsep}{0em}
    \setlength{\parskip}{0em}
}
{
  \end{oldthebibliography}
}

\title{Heavy quark diffusion coefficient with gradient flow}
\ShortTitle{Heavy quark diffusion coefficient with gradient flow}

\author*[a]{Viljami Leino}
\author[b,c,d]{Nora Brambilla}
\author[b,d]{Julian Mayer-Steudte}
\author[e]{Peter~Petreczky} 

\affiliation[a]{Helmholtz Institut Mainz, Johannes Gutenberg-Universität Mainz, D-55099 Mainz, Germany} 
\affiliation[b]{Physik Department, Technische Universit\"at M\"unchen,\\
James-Franck-Strasse 1, 85748 Garching, Germany}
\affiliation[c]{Institute for Advanced Study, Technische Universit\"at M\"unchen,\\
Lichtenbergstrasse 2 a, 85748 Garching, Germany}
\affiliation[d]{Munich Data Science Institute, Technische Universit\"at M\"unchen, \\
Walther-von-Dyck-Strasse 10, 85748 Garching, Germany}
\affiliation[e]{Physics Department, Brookhaven National Laboratory,
             Upton, New York 11973, USA}

\emailAdd{viljami.leino@tum.de}
\emailAdd{nora.brambilla@ph.tum.de}
\emailAdd{julian.mayer-steudte@tum.de}
\emailAdd{petreczk@bnl.gov}

\abstract{The heavy quark diffusion coefficient is encoded in the spectral functions of the chromo-electric and the chromo-magnetic correlators, of which the latter describes the T/M contribution. We study these correlators at two different temperatures $T=1.5\Tc$ and $T=10^4\Tc$ in the deconfined phase of SU(3) gauge theory. We use gradient flow for noise reduction. We perform both continuum and zero flow time limits to extract the heavy quark diffusion coefficient. Our results imply that the mass suppressed effects in the heavy quark diffusion coefficient are 20\% for bottom quarks and 34\% for charm quark at $T=1.5\Tc$.}

\FullConference{%
  The 39th International Symposium on Lattice Field Theory (Lattice2022),\\
  8-13 August, 2022 \\
  Bonn, Germany 
}

\begin{document}
\maketitle

\section{Introduction}\label{sec:intro}
The behavior of a heavy quark moving in a strongly coupled quark gluon plasma 
can be described by a set of transport coefficients.
In particular, 
the equilibration time of the heavy quark
can be described by Langevin dynamics that depend on three
related transport coefficients~\cite{Moore:2004tg}: the heavy quark momentum diffusion coefficient $\kappa$, 
the heavy quark diffusion coefficient $D_\mathrm{s}$, and the drag coefficient $\eta$. 
In this paper we focus on the heavy quark momentum diffusion coefficient $\kappa$.
The heavy quark momentum diffusion coefficient is known 
in perturbation theory up to mass-dependent contributions 
at NLO accuracy~\cite{Moore:2004tg,Svetitsky:1987gq,Caron-Huot:2008dyw}.
However, this NLO correction is sizable~\cite{Caron-Huot:2008dyw}, which invites for non-perturbative studies at strong coupling.
We will label this leading term in the \TM expansion as $\kappaE$.
Moreover, the first mass-dependent contribution when expanding the diffusion coefficient with respect to \TM has been studied in~\cite{Laine:2021uzs,Bouttefeux:2020ycy}, and we will label it as $\kappaB$. It is sensitive to chromo-magnetic screening and therefore,
is not calculable in perturbation theory~\cite{Bouttefeux:2020ycy}.

For measuring $\kappaEB$ we take an approach laid out 
by effective field theory and relate the momentum diffusion coefficient to correlators of field strength tensor components.
This allows us to circumvent
the problematic transport peak that is often encountered in transport coefficient calculations.
The leading contribution in the \TM expansion $\kappaE$ 
is related to a correlator of two chromo-electric fields~\cite{Caron-Huot:2009ncn,Brambilla:2016wgg}, 
and the leading \TM correction is related to a correlator of two chromo-magnetic fields ~\cite{Bouttefeux:2020ycy}. 
The diffusion coefficients $\kappaEB$ are then defined as a $\omega \rightarrow 0$ limit of the associated
spectral functions $\rhoEB(\omega)$. 

The diffusion coefficients $\kappaEB$ have been studied
on the lattice within this approach in pure gauge 
theory~\cite{Meyer:2010tt,Banerjee:2011ra,Francis:2015daa,Brambilla:2020siz,Banerjee:2022uge,Banerjee:2022gen}, 
using the multilevel algorithm~\cite{Luscher:2001up}.
Recently, the gradient flow algorithm~\cite{Luscher:2009eq,Luscher:2010iy} 
has been found to be useful for this quantity due to its renormalization properties and because it can be extended to theories with dynamical fermions more easily than multilevel simulations.
The heavy quark diffusion coefficient $\kappaEB$ has been measured with gradient flow 
in Refs.~\cite{Altenkort:2020fgs,Altenkort:2021ntw,Mayer-Steudte:2021hei,Brambilla:2022xbd}.

In this proceedings, we condense the results of our recent paper~\cite{Brambilla:2022xbd} for both the chromo-electric and chromo-magnetic correlators on the lattice using the gradient flow algorithm and determine the diffusion coefficient components $\kappaE$ and $\kappaB$ from the respective reconstructed spectral functions.

\section{Chromo-electric and Chromo-magnetic correlators}\label{sec:correlator}
The heavy quark momentum diffusion coefficient at leading order and including the leading $T/M$ corrections can be written as~\cite{Bouttefeux:2020ycy}:
\be
\kappa = \kappaE + \frac{2}{3}\langle \mathbf{v}^2 \rangle \kappaB\,,
\ee
where $\kappaE$ and $\kappaB$ can be extracted from Euclidean correlation functions by inverting a spectral function $\rho(\omega)$ and taking the zero frequency limit:
\ba
\GEB(\tau) &= \int_0^\infty \frac{\dd{\omega}}{\pi}\rhoEB(\omega,T)K(\omega,\tauT)\,,
\quad
K(\omega,\tauT) = 
\frac{\cosh\left(\frac{\omega}{T}\left(\tauT-\frac{1}{2}\right)\right)}{\sinh\left(\frac{\omega}{2T}\right)}\,, 
\label{eq:gefromrho} \\
\kappaEB &\equiv \lim_{\omega\rightarrow 0} \frac{2T\rhoEB(\omega,T)}{\omega}\label{eq:kappalimd}\,.
\ea
Here, in the heavy quark limit $M\gg \pi T$, 
the Euclidean correlators can be expressed in terms of  chromo-electric $E_i$ 
and chromo-magnetic $B_i$ fields~\cite{Caron-Huot:2009ncn,Casalderrey-Solana:2006fio,Bouttefeux:2020ycy}:
\ba
\GE(\tau) &= -\sum_{i=1}^{3} 
 \frac{\left\langle \ReTr\left[U(1/T,\tau)E_i(\tau,\mathbf{0})U(\tau,0)E_i(0,\mathbf{0})\right]\right\rangle}{3\left\langle\ReTr U(1/T,0)\right\rangle}\,,\label{eq:gelat} \\
\GB(\tau) &= \sum_{i=1}^{3} 
 \frac{\left\langle \ReTr\left[U(1/T,\tau)B_i(\tau,\mathbf{0})U(\tau,0)B_i(0,\mathbf{0})\right]\right\rangle}{3\left\langle\ReTr U(1/T,0)\right\rangle}\,,\label{eq:gblat}
\ea
where $T$ is the temperature and  $U(\tau_1,\tau_2)$ is a Wilson line in the Euclidean time direction.

In order to measure the Euclidean correlators we rely on the gradient flow algorithm~\cite{Luscher:2009eq,Luscher:2010iy}, that
systematically cools off the UV physics and automatically renormalizes the gauge invariant observables~\cite{Luscher:2011bx}.
This renormalization property of the gradient flow is especially useful for the correlators $\GEB$, which otherwise require an multiplicative renormalization on the lattice. 
However, for chromo-magnetic fields there is further renormalization required
both on the lattice and in continuum~\cite{Laine:2021uzs} that will not be automatically renormalized by the gradient flow. 
Moreover, since the gradient flow introduces a length scale $\sqrt{8\tauf}$, 
we have to make sure that the measurements at the length scale of interest $\tau$, which describes the separation between the chromo-electric or -magnetic fields, will not be affected by this new scale.
To avoid the mixing the scales, 
we will restrict our analysis to a regime $a/\tau \le \sqrt{8\tauf}/\tau \le 1/3$.

We have generated  
a set of pure-gauge SU(3) configurations using the standard Wilson gauge action at
two temperatures:
a low temperature $1.5\Tc$, and a high temperature $10^4\Tc$, with $\Tc$ being the deconfinement temperature.
The  temperatures are  related to the lattice spacing $a$ via the scale setting done in Ref.~\cite{Francis:2015lha}.
We consider lattices with varying numbers of temporal sites, $\Nt=20$,~24,~28, and~34,
and with corresponding spatial extents of $\Ns=48$,~48,~56, and~68 sites. 

\begin{figure}
    \includegraphics[width=0.5\textwidth]{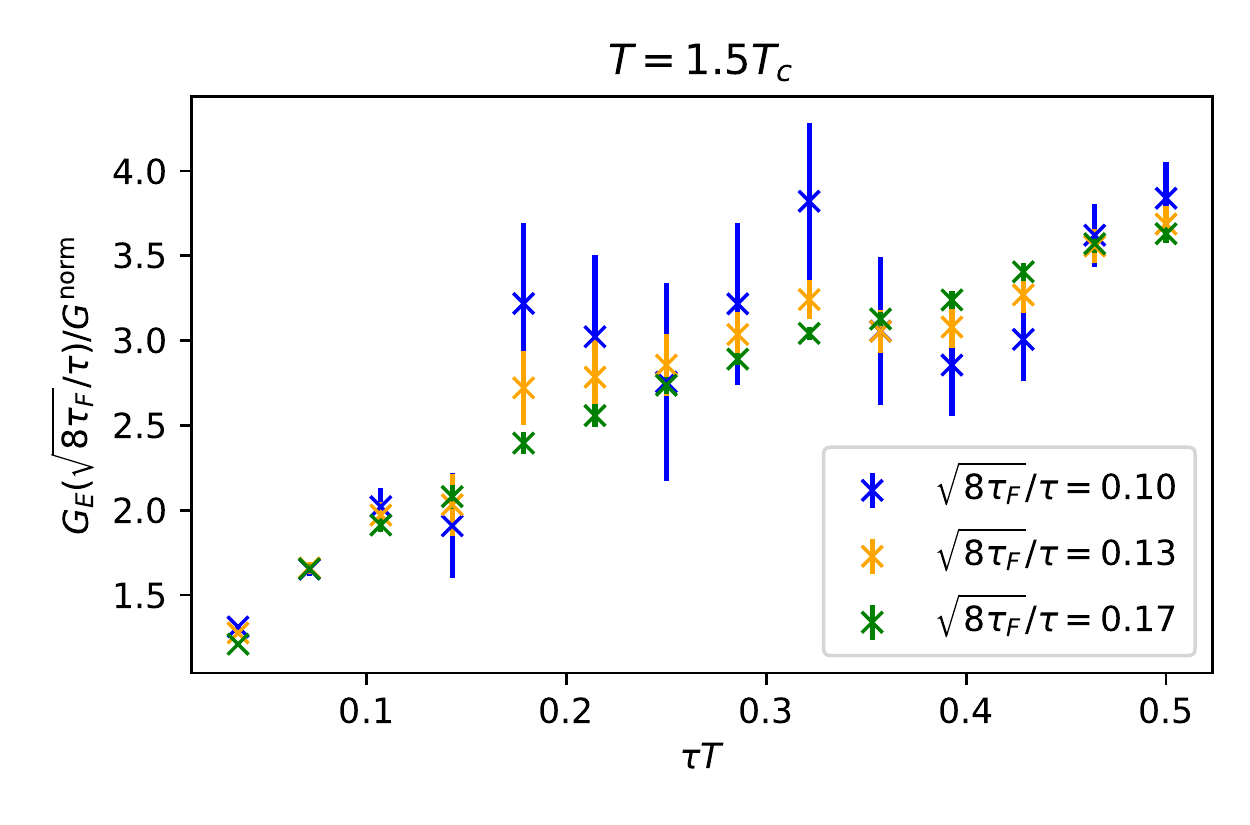}
    \includegraphics[width=0.5\textwidth]{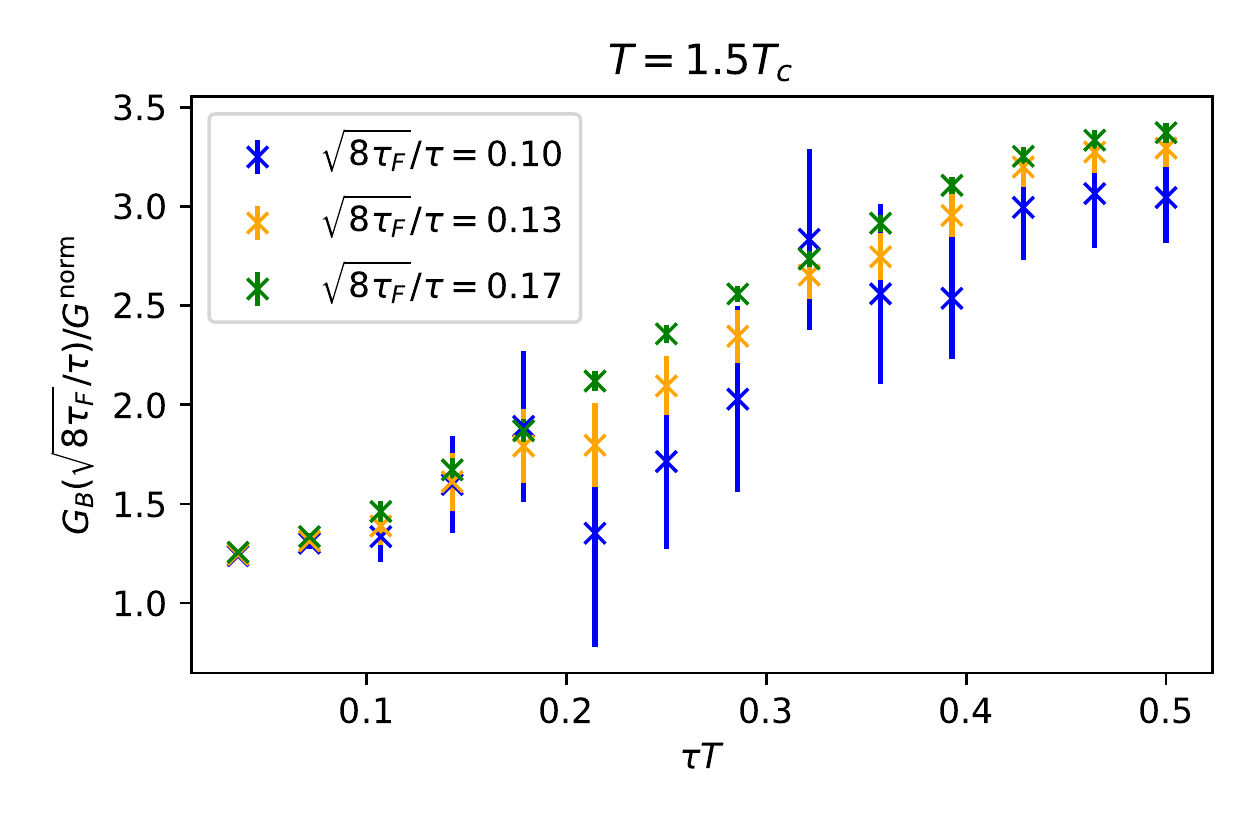}
    \caption{The normalized correlators $\GE$ (Left) and $\GB$ (Right) at fixed flow-time ratios, $\sqrt{8 \tauF}/\tau$ for the $N_t=28$ lattice at $T=1.5T_c$. 
    }
    \label{fig:corr_fixed_ratio_examples}
\end{figure}

We present the raw lattice measurements of $\GEB$ in Fig.~\ref{fig:corr_fixed_ratio_examples}.
Here we are using tree-level improved distances, such that we have scaled the correlators with the ratio of LO 
perturbative expressions of lattice and continuum perturbation theories. 
Furthermore, we normalize the data with the LO behavior:
\be\label{eq:gepert}
\frac{\GE^\mathrm{LO}(\tau)}{g^2\CF} \equiv G^{\rm norm}(\tau) = \pi^2 T^4 \left[\frac{\cos^2(\pi\tauT)}{\sin^4(\pi \tauT)}+\frac{1}{3\sin^2(\pi\tauT)}\right]\,.
\ee
We observe the statistical errors decreasing with increasing ratio $\sfrac{\sqrt{8\tauf}}{\tau}$
and that both correlators approach a common shape with increasing flow time.

\begin{figure}
    \includegraphics[width=0.5\textwidth]{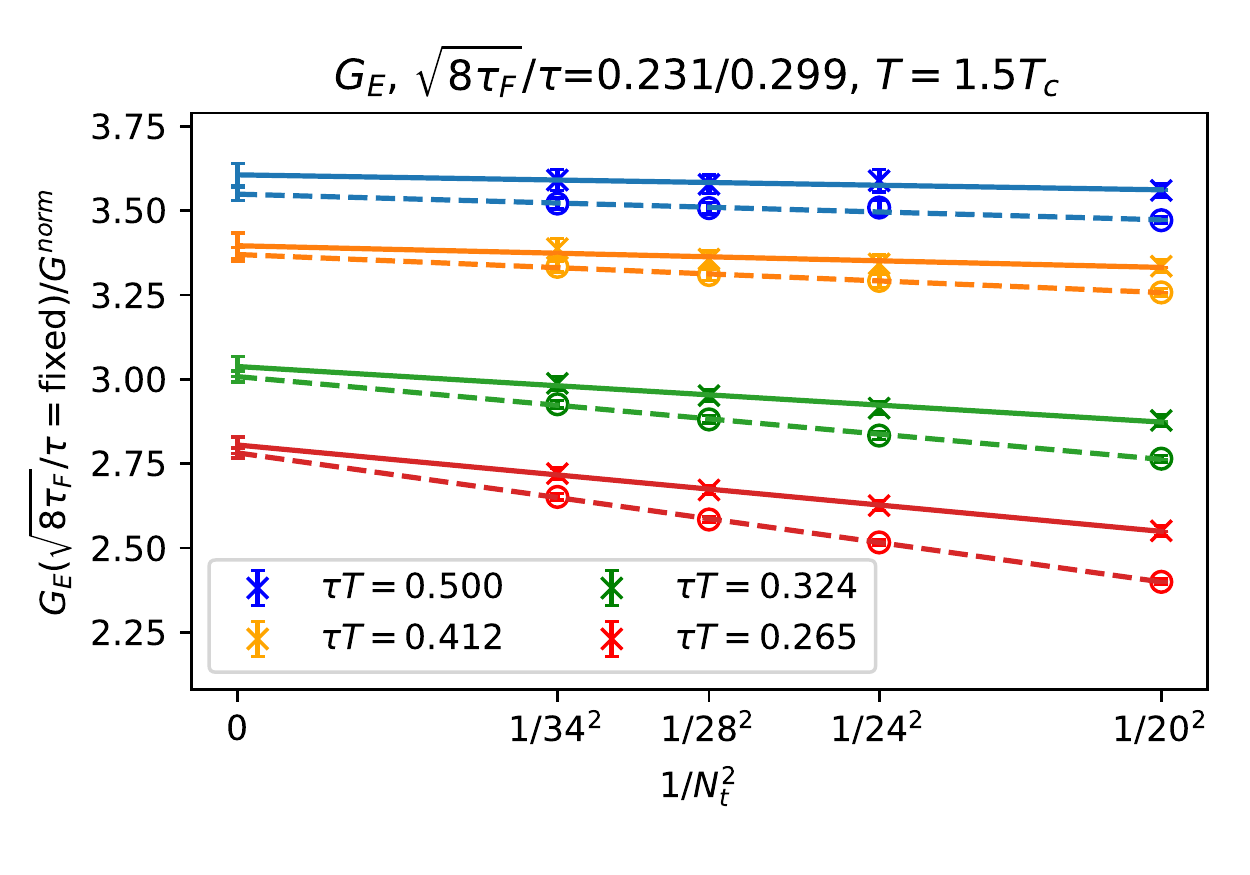}
    \includegraphics[width=0.5\textwidth]{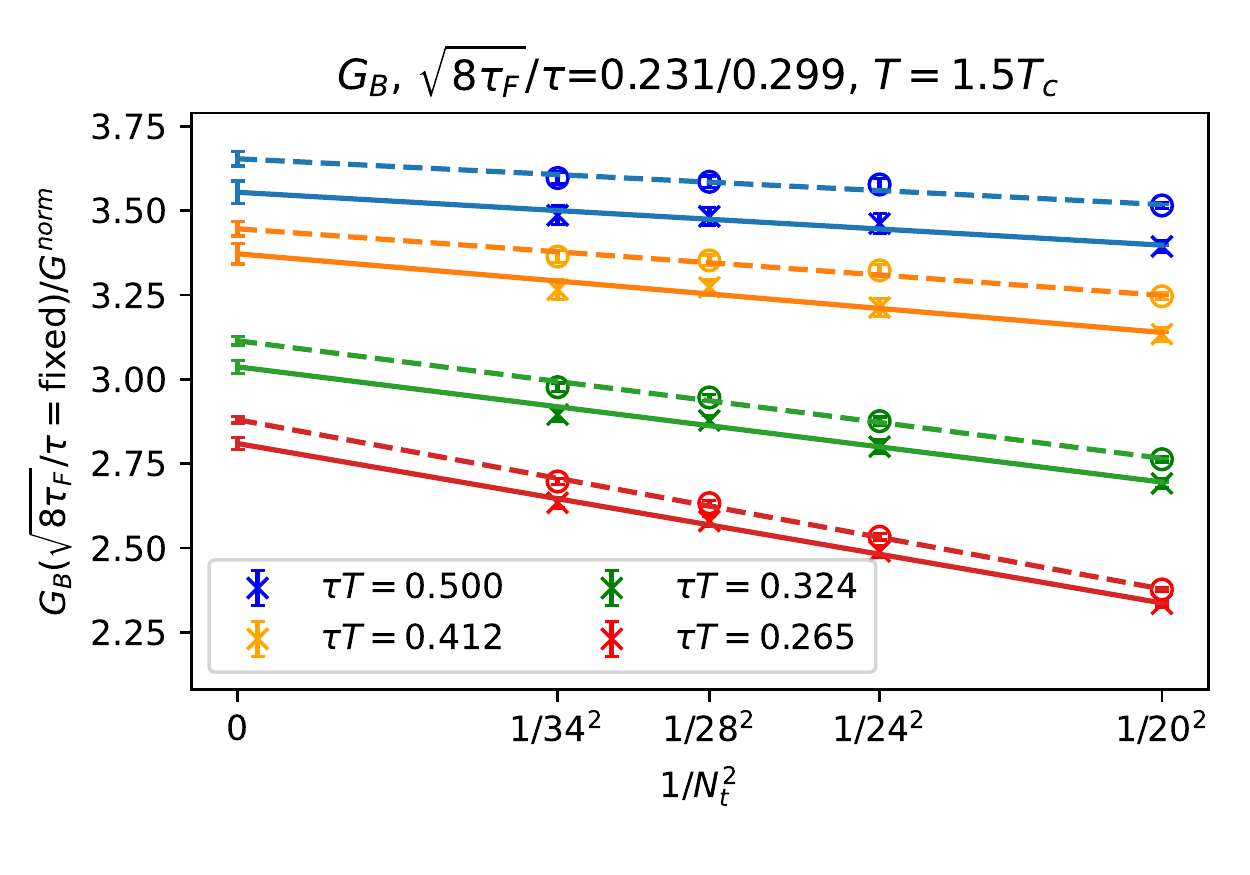}
    \caption{Examples of continuum extrapolations at fixed $\sqrt{8\tauf}/\tau$ for $\GE$ (Left) and $\GB$ (Right) correlators at $T=1.5T_c$. The dashed lines and circles present the limit taken at the lower edge of the flow time ratio of interest $\sfrac{\sqrt{8\tauf}}{\tau}=0.231$ while solid lines and asterisks have higher ratio of $\sfrac{\sqrt{8\tauf}}{\tau}=0.299$. 
    }
    \label{fig:cont_limit_example_limit}
\end{figure}
The raw lattice data at finite lattice spacing needs to be extrapolated to the continuum limit.
We start by interpolating the Euclidean correlator data for each lattice in $\tauT$ at fixed flow time ratio with cubic splines. 
Next, we perform a linear extrapolation in $1/\Nt^2 = (aT)^2$ of the correlators at the fixed interpolated $\tauT$, and fixed flow-time ratio positions, using all our lattice volumes for large separations
$\tauT>0.25$. For small separations $\tauT<0.25$, we drop the $\Nt=20$ lattice from the extrapolation. 
Based on our previous study~\cite{Brambilla:2020siz}, we do not expect there to be a notable dependence on the spatial size of the lattice. 
As an example, we show the continuum extrapolations at different $\tauT$ and $\sfrac{\sqrt{8\tauf}}{\tau}$ 
in Fig.~\ref{fig:cont_limit_example_limit},
where the continuum limits is shown at the edges of the $\sfrac{\sqrt{8\tauf}}{\tau}$ range, that we will later perform the zero flow time limit in.
\begin{figure}
    \includegraphics[width=0.5\textwidth]{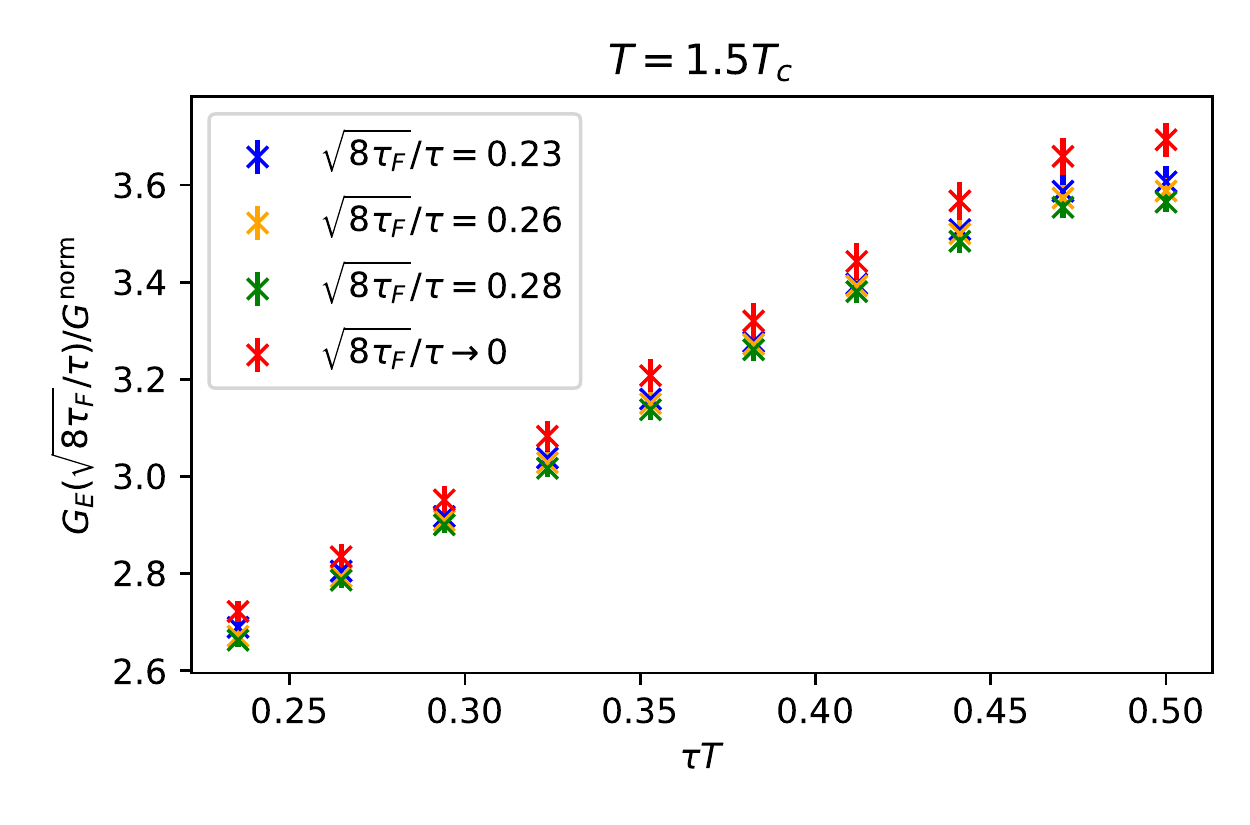}
    \includegraphics[width=0.5\textwidth]{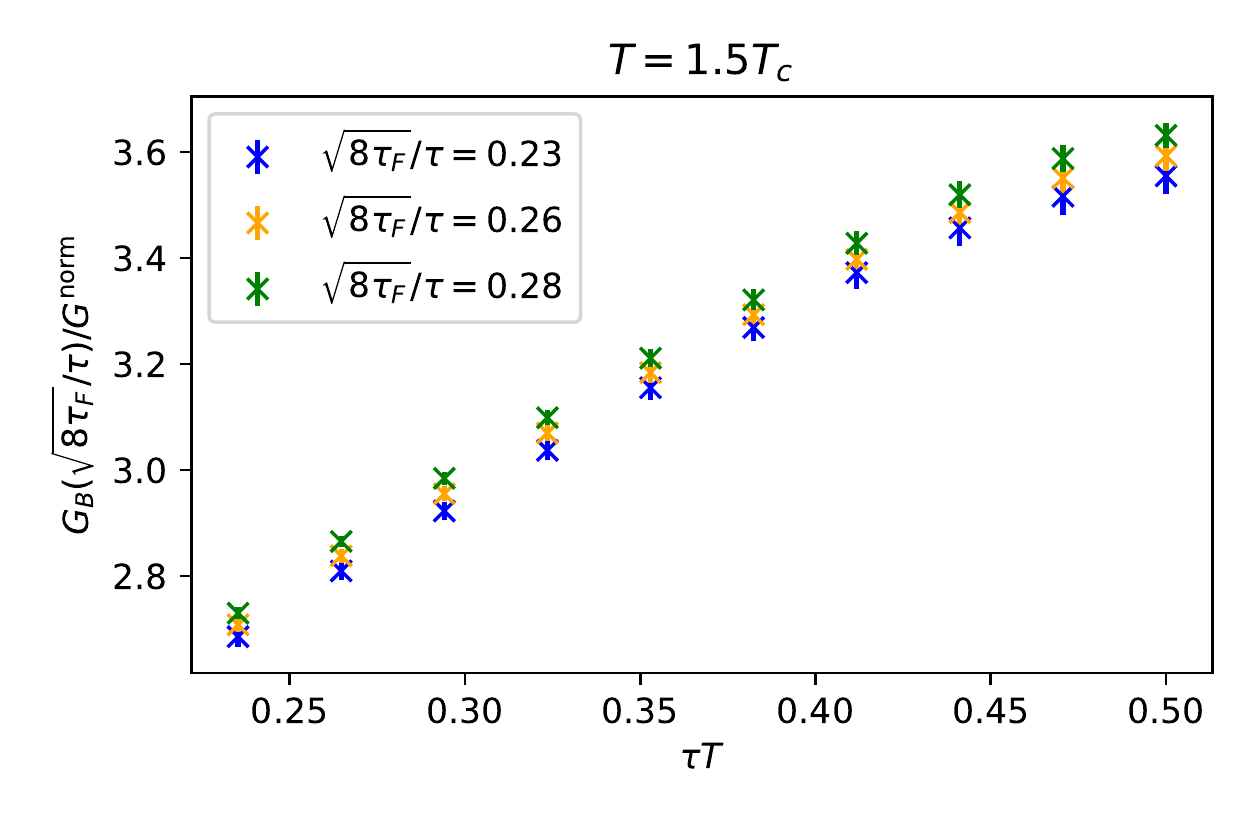}
    \caption{The continuum limits of $\GE$ (Left) and $\GB$ (Right) 
             at $T=1.5\Tc$ for different fixed flow-time ratios.
    }
    \label{fig:cont_limit_example}
\end{figure}
Finally in Fig.~\ref{fig:cont_limit_example},
we show the final continuum limit of $\GE$ and $\GB$.

Once the discretization effects have been removed by the continuum limit, 
we will need to get rid of distortions due to gradient flow.
For this we extrapolate the lattice results of $\GE$ to zero flow time using linear ansatz.
\begin{figure}
    \includegraphics[width=0.5\textwidth]{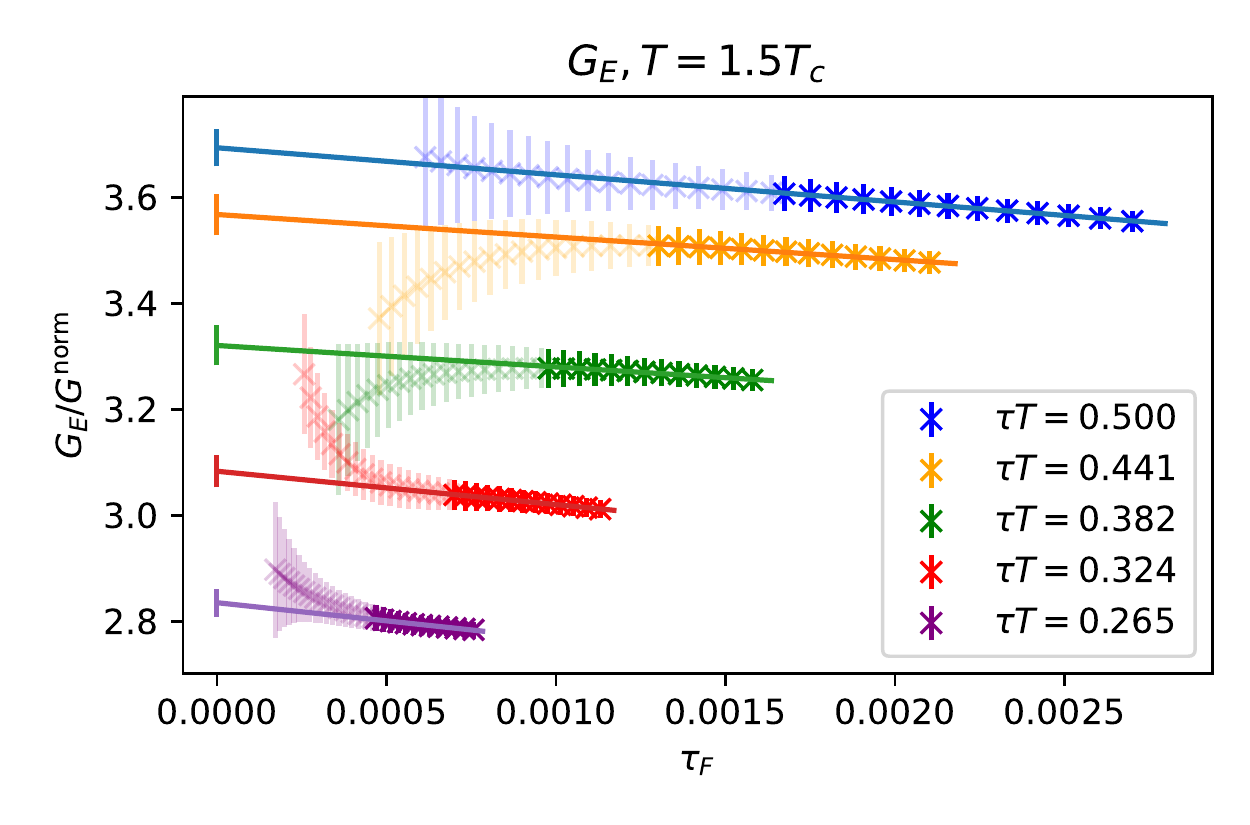}
    \includegraphics[width=0.5\textwidth]{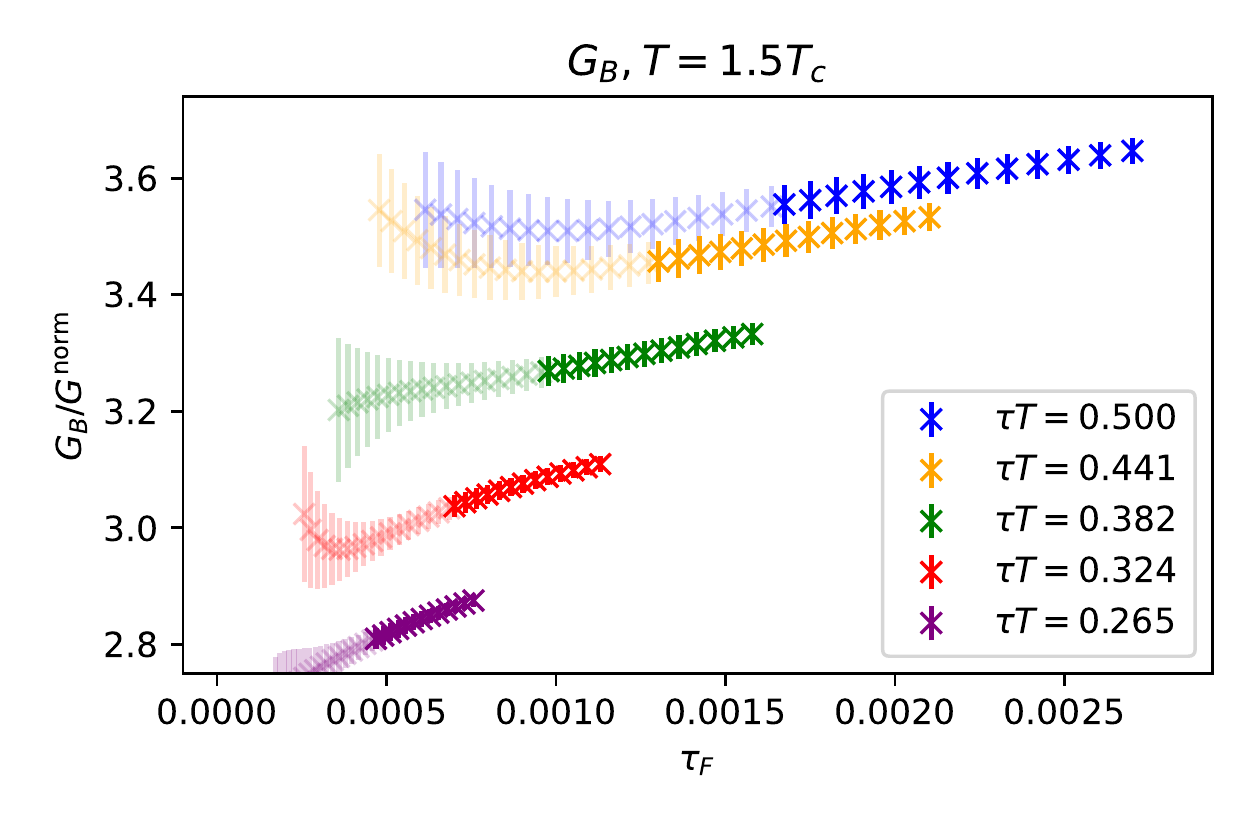}
    \caption{The flow time dependence of $\GE$ (Left) and $\GB$ (Right) at 
             $T=1.5\Tc$ in the continuum limit.
             }
    \label{fig:Ge_flowed}
\end{figure}
We show representative cases of these extrapolations on the left side of Fig.~\ref{fig:Ge_flowed}.
Meanwhile, the flow time dependence of $\GB$ is shown on the right side of Fig.~\ref{fig:Ge_flowed} 
and it appears to be quite different from the flow time dependence of the chromo-electric correlator. 
The flow time dependence of $\GB$ appears to be roughly linear but has a slope with opposite sign.
This difference is expected and is due to 
the non-trivial renormalization of the $\GB$, which will be discussed in the next section.
Due to finite anomalous dimension in the chromo-magnetic correlator, we refrain from taking 
the zero flow time limit of $\GB$ and instead will extract $\kappaB$ at finite flow time.

\section{Measuring the diffusion coefficient on the lattice}\label{sec:GEkappa}
Similarly to our preceding multilevel study~\cite{Brambilla:2020siz},
we extract $\kappaE$ and $\kappaB$ using Eq.\eqref{eq:gefromrho} and modeling the spectral function $\rho(\omega)$ using the UV and IR behaviors motivated by the perturbation theory~\cite{Burnier:2010rp,Banerjee:2022uge}:
\be
\rhoEB^\mathrm{IR}(\omega,T) = \frac{\omega\kappa}{2T}\label{eq:rhoIR}\,, \quad\text{and}\quad
\rhoEB^\mathrm{UV}(\omega,T) = \frac{g^2(\mu_\omega^\mathrm{opt})\CF\omega^3}{6\pi}\,,
\ee
where the scale of the running coupling $\mu_\omega$ 
has been chosen so that the NLO contribution to UV spectral function vanishes. 
For $\rhoE$, this scale is easy to determine\cite{Burnier:2010rp}:
\be\label{muom1}
\ln(\mu_\omega^\mathrm{opt})=\ln(2\omega) +
\frac{(24\pi^2-149)}{66}\,.
\ee
For the magnetic spectral function $\rhoB$, 
the situation is more complicated due to required renormalization~\cite{Bouttefeux:2020ycy,Laine:2021uzs}.
In order to study the chromo-magnetic correlator $\GB$ at the zero flow time,
we use the relation for the UV part of $G_B$ at non-zero
flow time to the corresponding renormalized correlator in $\MSb$ scheme:
\be
\GB^\mathrm{flow, UV}(\tau,\tauF)=(1+\gamma_0 g^2 \ln(\mu \sqrt{8 \tauF}))^2 Z_\mathrm{flow} \GB^{\MSb,\mathrm{UV}}(\tau,\mu) 
+h_0 \cdot(\tauF/\tau)\,,\label{eq:gbgfmsb}
\ee
where $h_0$ is a constant and $\gamma_0=3/(8\pi^2)$ 
is the anomalous dimension of the chromo-magnetic field~\cite{Banerjee:2022uge}. 
Using the NLO result from Ref.~\cite{Banerjee:2022uge} and neglecting the distortions due to finite flow
time by setting $h_0$ to zero, we can arrive to a scale:
\be\label{muom3}
\mu^\mathrm{opt}_\omega=(\sqrt{A} \omega)^{1-\gamma_0/\beta_0} \cdot (8 \tauF)^{-\gamma_0/(2 \beta_0)}\,,
\quad
A = \exp\left[\frac{134}{35} - \frac{8\pi^2}{5} - \ln 4\right]\,,
\ee
which will give correct $\rho^\mathrm{UV}$ up to a multiplicative constant $Z_\mathrm{flow}$.

From our previous study~\cite{Brambilla:2020siz},
we know that the NLO behavior is not quite enough to capture the zero temperature part of the Euclidean correlators, hence an additional normalization constant $C_n$ needs to be introduced as a fit parameter.
For chromo-magnetic correlators, we will absorb the unknown constant $Z_\mathrm{flow}$ into $C_n$.
Furthermore, 
we assume that $\rhoEB(\omega,T)=\rho^\mathrm{IR}(\omega,T)$ for $\omega<\omega^\mathrm{IR}$
and $\rhoEB(\omega,T)=\rhoEB^\mathrm{UV}(\omega,T)$ for $\omega>\omega^\mathrm{UV}$,
where $\omega^\mathrm{IR}$ and $\omega^\mathrm{UV}$ are the limiting values of $\omega$ for which we can trust the above behaviors.
In the region $\omega^\mathrm{IR}<\omega<\omega^\mathrm{UV}$, the form of the spectral function is generally not known.
Hence, for a given value of $\kappaEB$, we construct the model spectral function that is given by $\rhoEB^\mathrm{UV}$ in $\omega > \omega^\mathrm{UV}$,
$\rhoEB^\mathrm{IR}$ in $\omega<\omega^\mathrm{IR}$, and vary various forms of $\rhoEB(\omega)$ for the intermediate $\omega^\mathrm{IR} \le \omega \le \omega^\mathrm{UV}$
such that the total spectral function is continuous. 
For the exact functional forms, we refer the reader to our main paper~\cite{Brambilla:2022xbd}.

\begin{figure}
\centering
\resizebox{0.6\textwidth}{!}{\input{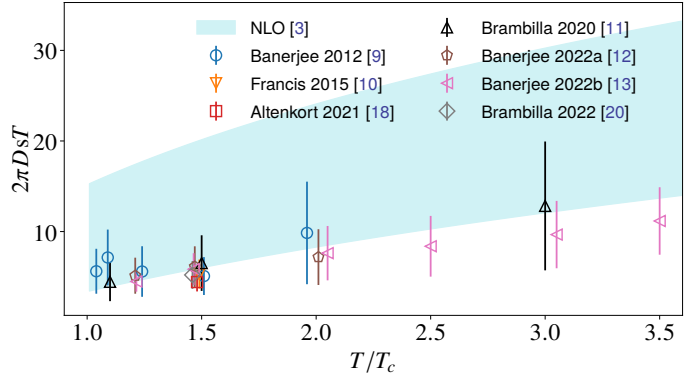}}
\caption{Our result of $\kappaE$ (Grey diamond) compared to existing lattice results. Points that are at the same temperature, have been slightly shifted horizontally for better visibility.}
\label{fig:Dscompare}
\end{figure}
We now hold all the information needed to extract $\kappaE$ and $\kappaB$. 
The extraction of $\kappaE$ then proceeds as follows. 
We take the continuum limit data at the zero flow time limit 
and perform a least squares fit to Eq.~\eqref{eq:gefromrho} 
with our set of different spectral function models.
In order to estimate systematics, we vary the range of points included in the fits by dropping certain
amount of early points $\tauT_\mathrm{min}$, we vary the scale with a factor of two, and we use multiple different spectral function models.
For $\kappaE$ we then get as a final result:
\be
\label{eq:kappaE_1.5Tc}
1.70 \le \frac{\kappaE}{T^3} \le 3.12\,,\;\text{at}\;T=1.5\Tc\,,\quad\text{and}\quad
0.02 \le \frac{\kappaE}{T^3} \le 0.16\,,\;\text{at}\;T=10^4\Tc\,. 
\ee
The result for $T=1.5\Tc$ is in full agreement with the existing results. 
We show this in Fig.~\ref{fig:Dscompare} for the spatial diffusion coefficient 
$D_\mathrm{s}=\sfrac{2T^2}{\kappa}$.
For $T=10^4\Tc$, we are in agreement 
with our previous result $0<\sfrac{\kappaE}{T^3}<0.1$~\cite{Brambilla:2020siz}. 
The new result has a slightly larger errors due to gradient flow analysis having more strict fit regimes. 
However, we can for the first time observe a nonzero minimum for $\sfrac{\kappaE}{T^3}$ 
at very large temperature. 

\begin{figure}
    \includegraphics[width=0.5\textwidth]{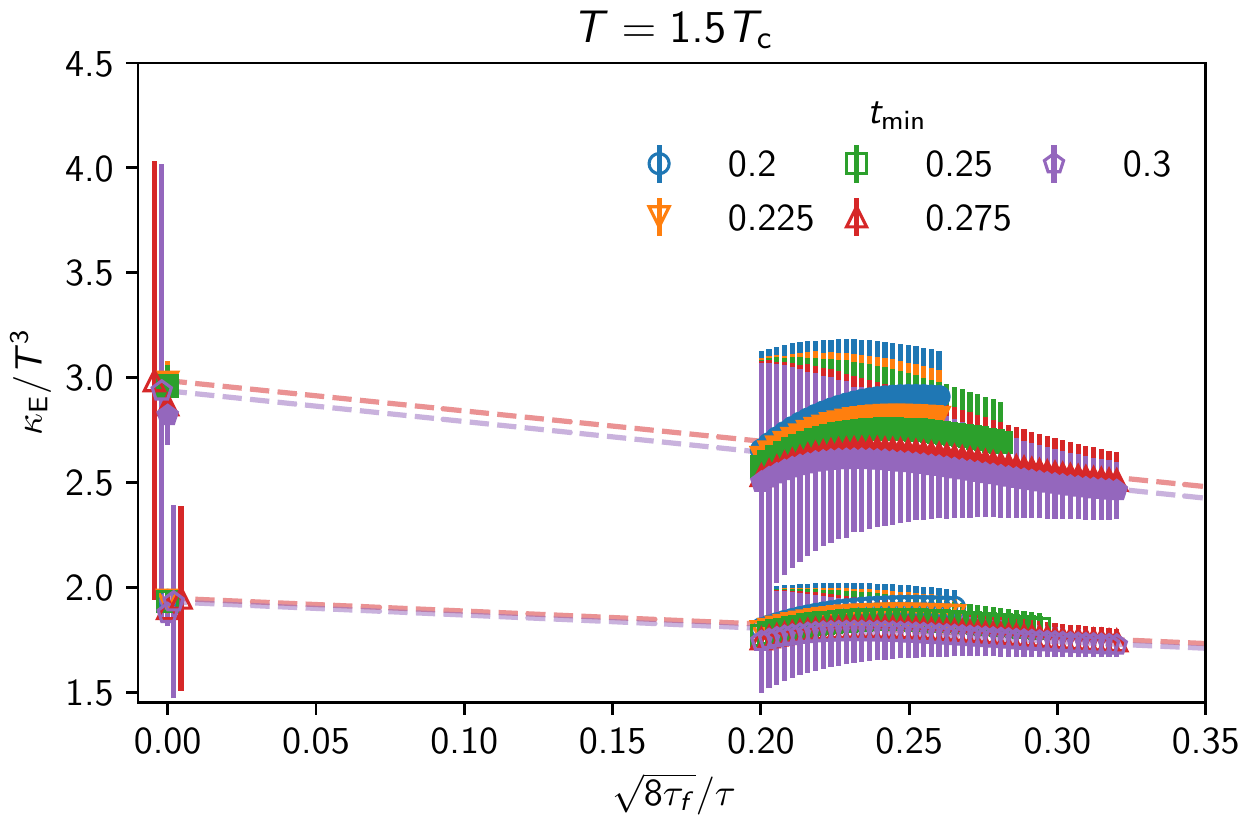}
    \includegraphics[width=0.5\textwidth]{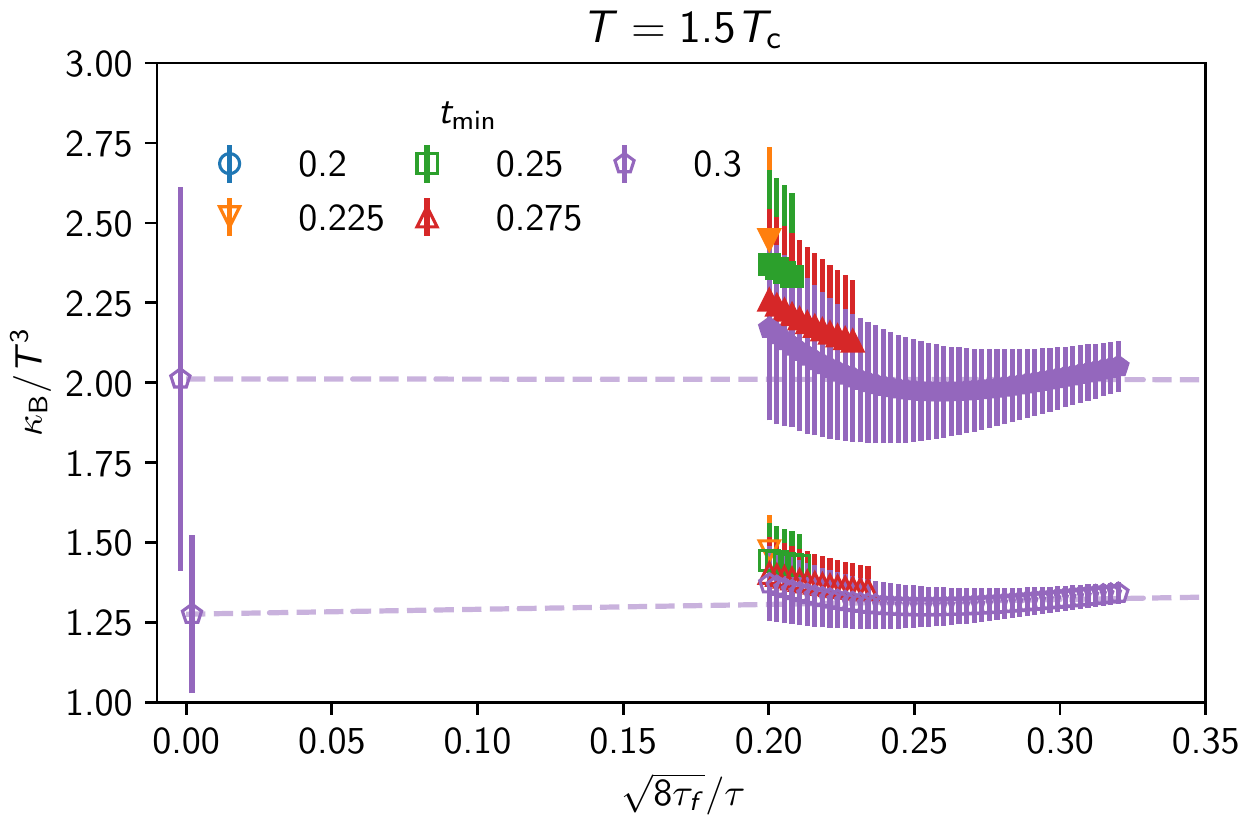}
    \caption{The heavy quark momentum diffusion coefficients $\kappaE/T^3$ (Left) and $\kappaB/T^3$ (Right)
             at different flow time ratios $\sqrt{8\tauf}/\tau$. The filled/unfilled points depict the 
             different models used for $\rho(\omega)$, 
             while different symbols show the different $\tauT_\mathrm{min}$. 
             }
    \label{fig:Ekappaatfiniteratio}
\end{figure}
For $\kappaB$ we had to invert the spectral function at a finite flow time, which means we need to do the zero flow time extrapolation at the level of $\kappaB$. To further reinforce this method, we show on the left side of Fig.~\ref{fig:Ekappaatfiniteratio} that such limit also works for $\kappaE$. Then on the right hand side of Fig~\ref{fig:Ekappaatfiniteratio}, we show the flow time dependence of $\kappaB$. We observe that by using flow time dependent scale, at sufficiently large $\tauT_\mathrm{min}$, there is very little flow time dependence left.
If we take the total variation at finite flow time to be the error of $\kappaB$, we get for 
$T=1.5\Tc$: $1.23< \sfrac{\kappaB}{T^3} < 2.74$. 
We can then proceed to take the zero flow time limit in the linear regime $\sqrt{8\tauF}/\tau \ge 0.25$, similar to what we learned to work with in the case of $\GE$. We get at the zero flow time limit
the final result for $\kappaB$:
\be
\label{eq:kappaB_1.5Tc}
1.03 \le \frac{\kappaB}{T^3} \le 2.61\,.  
\ee
This result is well in agreement with the recent result~\cite{Banerjee:2022uge}, 
that got $1.0 \le \kappaB/T^3 \le 2.1$. 
The current data is not accurate enough to determine $\kappaB$ at $T=10^4 T_c$.

We can now see what happens if we try to combine $\kappaE$ and $\kappaB$.
Using the lattice results for $\langle \mathbf{v}^2 \rangle$
from~\cite{Petreczky:2008px} 
for charm and bottom quarks $\langle \mathbf{v}^2 \rangle_\mathrm{charm} \simeq 0.51$
and $\langle \mathbf{v}^2 \rangle_\mathrm{bottom}\simeq 0.3$, 
we estimate that the mass suppressed effect on
the heavy quark diffusion coefficient is 34\% and 20\% for charm and bottom quark, respectively.

\FloatBarrier
\acknowledgments{
The lattice QCD calculations have been performed using the publicly available \href{https://web.physics.utah.edu/~detar/milc/milcv7.html}{MILC code}. The simulations were carried out on the computing facilities of the Computational Center for Particle and Astrophysics (C2PAP) in the project Calculation of finite T QCD correlators (pr83pu) and of the SuperMUC cluster at the Leibniz-Rechenzentrum (LRZ) in the project The role of the charm-quark for the QCD coupling constant (pn56bo). This research was funded by the Deutsche Forschungsgemeinschaft (DFG, German Research Foundation) cluster of excellence “ORIGINS” (\href{www.origins-cluster.de}{www.origins-cluster.de}) under Germany’s Excellence Strategy EXC-2094-390783311.
}

\bibliographystyle{jhep_modified}
\bibliography{kappa.bib}

\providecommand{\href}[2]{#2}\begingroup\raggedright\begin{thebibliography}{10}

\bibitem{Moore:2004tg}
G.~D. Moore and D.~Teaney, \emph{{How much do heavy quarks thermalize in a
  heavy ion collision?}},
  \href{https://doi.org/10.1103/PhysRevC.71.064904}{\emph{Phys. Rev.}
  {\bfseries C71} (2005) 064904}
  [\href{https://arxiv.org/abs/hep-ph/0412346}{{\ttfamily hep-ph/0412346}}].
\bibitem{Svetitsky:1987gq}
B.~Svetitsky, \emph{{Diffusion of charmed quarks in the quark-gluon plasma}},
  \href{https://doi.org/10.1103/PhysRevD.37.2484}{\emph{Phys. Rev.} {\bfseries
  D37} (1988) 2484}.
\bibitem{Caron-Huot:2008dyw}
S.~Caron-Huot and G.~D. Moore, \emph{{Heavy quark diffusion in QCD and N=4 SYM
  at next-to-leading order}},
  \href{https://doi.org/10.1088/1126-6708/2008/02/081}{\emph{JHEP} {\bfseries
  02} (2008) 081} [\href{https://arxiv.org/abs/0801.2173}{{\ttfamily
  0801.2173}}].
\bibitem{Laine:2021uzs}
M.~Laine, \emph{{1-loop matching of a thermal Lorentz force}},
  \href{https://doi.org/10.1007/JHEP06(2021)139}{\emph{JHEP} {\bfseries 06}
  (2021) 139} [\href{https://arxiv.org/abs/2103.14270}{{\ttfamily
  2103.14270}}].
\bibitem{Bouttefeux:2020ycy}
A.~Bouttefeux and M.~Laine, \emph{{Mass-suppressed effects in heavy quark
  diffusion}}, \href{https://doi.org/10.1007/JHEP12(2020)150}{\emph{JHEP}
  {\bfseries 12} (2020) 150}
  [\href{https://arxiv.org/abs/2010.07316}{{\ttfamily 2010.07316}}].
\bibitem{Caron-Huot:2009ncn}
S.~Caron-Huot, M.~Laine and G.~D. Moore, \emph{{A Way to estimate the heavy
  quark thermalization rate from the lattice}},
  \href{https://doi.org/10.1088/1126-6708/2009/04/053}{\emph{JHEP} {\bfseries
  04} (2009) 053} [\href{https://arxiv.org/abs/0901.1195}{{\ttfamily
  0901.1195}}].
\bibitem{Brambilla:2016wgg}
N.~Brambilla, M.~A. Escobedo, J.~Soto and A.~Vairo, \emph{{Quarkonium
  suppression in heavy-ion collisions: an open quantum system approach}},
  \href{https://doi.org/10.1103/PhysRevD.96.034021}{\emph{Phys. Rev.}
  {\bfseries D96} (2017) 034021}
  [\href{https://arxiv.org/abs/1612.07248}{{\ttfamily 1612.07248}}].
\bibitem{Meyer:2010tt}
H.~B. Meyer, \emph{{The errant life of a heavy quark in the quark-gluon
  plasma}}, \href{https://doi.org/10.1088/1367-2630/13/3/035008}{\emph{New J.
  Phys.} {\bfseries 13} (2011) 035008}
  [\href{https://arxiv.org/abs/1012.0234}{{\ttfamily 1012.0234}}].
\bibitem{Banerjee:2011ra}
D.~Banerjee, S.~Datta, R.~Gavai and P.~Majumdar, \emph{{Heavy Quark Momentum
  Diffusion Coefficient from Lattice QCD}},
  \href{https://doi.org/10.1103/PhysRevD.85.014510}{\emph{Phys. Rev.}
  {\bfseries D85} (2012) 014510}
  [\href{https://arxiv.org/abs/1109.5738}{{\ttfamily 1109.5738}}].
\bibitem{Francis:2015daa}
A.~Francis, O.~Kaczmarek, M.~Laine, T.~Neuhaus and H.~Ohno,
  \emph{{Nonperturbative estimate of the heavy quark momentum diffusion
  coefficient}}, \href{https://doi.org/10.1103/PhysRevD.92.116003}{\emph{Phys.
  Rev.} {\bfseries D92} (2015) 116003}
  [\href{https://arxiv.org/abs/1508.04543}{{\ttfamily 1508.04543}}].
\bibitem{Brambilla:2020siz}
N.~Brambilla, V.~Leino, P.~Petreczky and A.~Vairo, \emph{{Lattice QCD
  constraints on the heavy quark diffusion coefficient}},
  \href{https://doi.org/10.1103/PhysRevD.102.074503}{\emph{Phys. Rev. D}
  {\bfseries 102} (2020) 074503}
  [\href{https://arxiv.org/abs/2007.10078}{{\ttfamily 2007.10078}}].
\bibitem{Banerjee:2022uge}
D.~Banerjee, S.~Datta and M.~Laine, \emph{{Lattice study of a magnetic
  contribution to heavy quark momentum diffusion}},
  \href{https://doi.org/10.1007/JHEP08(2022)128}{\emph{JHEP} {\bfseries 08}
  (2022) 128} [\href{https://arxiv.org/abs/2204.14075}{{\ttfamily
  2204.14075}}].
\bibitem{Banerjee:2022gen}
D.~Banerjee, R.~Gavai, S.~Datta and P.~Majumdar, \emph{{Temperature dependence
  of the static quark diffusion coefficient}},
  \href{https://arxiv.org/abs/2206.15471}{{\ttfamily 2206.15471}}.
\bibitem{Luscher:2001up}
M.~Lüscher and P.~Weisz, \emph{{Locality and exponential error reduction in
  numerical lattice gauge theory}},
  \href{https://doi.org/10.1088/1126-6708/2001/09/010}{\emph{JHEP} {\bfseries
  09} (2001) 010} [\href{https://arxiv.org/abs/hep-lat/0108014}{{\ttfamily
  hep-lat/0108014}}].
\bibitem{Luscher:2009eq}
M.~Lüscher, \emph{{Trivializing maps, the Wilson flow and the HMC algorithm}},
  \href{https://doi.org/10.1007/s00220-009-0953-7}{\emph{Commun. Math. Phys.}
  {\bfseries 293} (2010) 899}
  [\href{https://arxiv.org/abs/0907.5491}{{\ttfamily 0907.5491}}].
\bibitem{Luscher:2010iy}
M.~Lüscher, \emph{{Properties and uses of the Wilson flow in lattice QCD}},
  \href{https://doi.org/10.1007/JHEP08(2010)071}{\emph{JHEP} {\bfseries 08}
  (2010) 071} [\href{https://arxiv.org/abs/1006.4518}{{\ttfamily 1006.4518}}].
\bibitem{Altenkort:2020fgs}
L.~Altenkort, A.~M. Eller, O.~Kaczmarek, L.~Mazur, G.~D. Moore and H.-T. Shu,
  \emph{{Heavy quark momentum diffusion from the lattice using gradient flow}},
  \href{https://doi.org/10.1103/PhysRevD.103.014511}{\emph{Phys. Rev. D}
  {\bfseries 103} (2021) 014511}
  [\href{https://arxiv.org/abs/2009.13553}{{\ttfamily 2009.13553}}].
\bibitem{Altenkort:2021ntw}
L.~Altenkort, A.~M. Eller, O.~Kaczmarek, L.~Mazur, G.~D. Moore and H.-T. Shu,
  \emph{{Continuum extrapolation of the gradient-flowed color-magnetic
  correlator at $1.5\,T_c$}},  in \emph{{38th International Symposium on
  Lattice Field Theory}}, 11, 2021,
  \href{https://arxiv.org/abs/2111.12462}{{\ttfamily 2111.12462}}.
\bibitem{Mayer-Steudte:2021hei}
J.~Mayer-Steudte, N.~Brambilla, V.~Leino and P.~Petreczky,
  \emph{{Chromoelectric and chromomagnetic correlators at high temperature from
  gradient flow}},  in \emph{{38th International Symposium on Lattice Field
  Theory}}, 11, 2021, \href{https://arxiv.org/abs/2111.10340}{{\ttfamily
  2111.10340}}.
\bibitem{Brambilla:2022xbd}
{\scshape TUMQCD} collaboration, N.~Brambilla, V.~Leino, J.~Mayer-Steudte and
  P.~Petreczky, \emph{{Heavy quark diffusion coefficient with gradient flow}},
  \href{https://arxiv.org/abs/2206.02861}{{\ttfamily 2206.02861}}.
\bibitem{Casalderrey-Solana:2006fio}
J.~Casalderrey-Solana and D.~Teaney, \emph{{Heavy quark diffusion in strongly
  coupled N=4 Yang-Mills}},
  \href{https://doi.org/10.1103/PhysRevD.74.085012}{\emph{Phys. Rev. D}
  {\bfseries 74} (2006) 085012}
  [\href{https://arxiv.org/abs/hep-ph/0605199}{{\ttfamily hep-ph/0605199}}].
\bibitem{Luscher:2011bx}
M.~Lüscher and P.~Weisz, \emph{{Perturbative analysis of the gradient flow in
  non-abelian gauge theories}},
  \href{https://doi.org/10.1007/JHEP02(2011)051}{\emph{JHEP} {\bfseries 02}
  (2011) 051} [\href{https://arxiv.org/abs/1101.0963}{{\ttfamily 1101.0963}}].
\bibitem{Francis:2015lha}
A.~Francis, O.~Kaczmarek, M.~Laine, T.~Neuhaus and H.~Ohno, \emph{{Critical
  point and scale setting in SU(3) plasma: An update}},
  \href{https://doi.org/10.1103/PhysRevD.91.096002}{\emph{Phys. Rev.}
  {\bfseries D91} (2015) 096002}
  [\href{https://arxiv.org/abs/1503.05652}{{\ttfamily 1503.05652}}].
\bibitem{Burnier:2010rp}
Y.~Burnier, M.~Laine, J.~Langelage and L.~Mether, \emph{{Colour-electric
  spectral function at next-to-leading order}},
  \href{https://doi.org/10.1007/JHEP08(2010)094}{\emph{JHEP} {\bfseries 08}
  (2010) 094} [\href{https://arxiv.org/abs/1006.0867}{{\ttfamily 1006.0867}}].
\bibitem{Petreczky:2008px}
P.~Petreczky, \emph{{On temperature dependence of quarkonium correlators}},
  \href{https://doi.org/10.1140/epjc/s10052-009-0942-1}{\emph{Eur. Phys. J. C}
  {\bfseries 62} (2009) 85} [\href{https://arxiv.org/abs/0810.0258}{{\ttfamily
  0810.0258}}].
\end{thebibliography}\endgroup

\end{document}